# Implementing NAT Hole Punching with QUIC

Jinyu Liang
College of Electronics
and Information Engineering
Shenzhen University
Shenzhen, China
2210434004@email.szu.edu.cn

Wei Xu
Computer Science Department
New York University
New York, USA
wx317@nyu.edu

Taotao Wang
College of Electronics
and Information Engineering
Shenzhen University
Shenzhen, China
ttwang@szu.edu.cn

Qing Yang
College of Electronics
and Information Engineering
Shenzhen University
Shenzhen, China
yang.qing@szu.edu.cn

Shengli Zhang
College of Electronics
and Information Engineering
Shenzhen University
Shenzhen, China
zsl@szu.edu.cn

*Abstract*—The widespread adoption of Network Address Translation (NAT) technology has led to a significant number of network end nodes being located in private networks behind NAT devices, impeding direct communication between these nodes. To solve this problem, a technique known as "hole punching" has been devised for NAT traversal to facilitate peer-to-peer communication among end nodes located in distinct private networks. However, as the increasing demands for speed and security in networks, TCP-based hole punching schemes gradually show performance drawbacks. Therefore, we present a QUIC-based hole punching scheme for NAT traversal. Through a comparative analysis of the hole punching time between QUIC-based and TCP-based protocols, we find that the QUIC-based scheme effectively reduces the hole punching time, exhibiting a pronounced advantage in weak network environments. Furthermore, in scenarios where the hole punched connection is disrupted due to factors such as network transitions or NAT timeouts, this paper evaluates two schemes for restoring the connection: QUIC connection migration and re-punching. Our results show that QUIC connection migration for connection restoration saves 2 RTTs compared to QUIC re-punching, and 3 RTTs compared to TCP re-punching, effectively reducing the computational resources consumption for re-punching.

*Keywords—QUIC, TCP, NAT Hole Punching, Connection Migration*

## I. Introduction

Network Address Translation (NAT) technology, introduced in 1994 by Egevang et al. [1], mitigates the scarcity of IPv4 addresses by mapping private IP addresses and ports to public IP addresses and ports. This enables devices within private networks to access the Internet using public addresses. The authors of [2] state that approximately 80% of end nodes in the Peer-to-Peer (P2P) streaming system PPLive are located within private networks behind NAT devices, obstructing direct P2P communication. Wang et al. note that a small subset of nodes within the Bitcoin network facilitates 89% of transaction propagation [3]. This phenomenon arises due to the majority of end nodes residing behind NAT devices and firewalls, rendering them challenging for other nodes to discover and interact with in transactions. Yin et al. point out that the presence of NAT devices may impede the effective discovery of some nodes in content distribution networks (CDNs), leading to content request concentration on a few reachable nodes, which impacts the overall performance and stability of the CDN network [4]. NAT alleviates the issue of IPv4 address scarcity to some extent, but it also increases the complexity of network management and limits the development of P2P communication.

To resolve the issue of P2P communication hindered by NAT, scholars have introduced a technique known as "hole punching," which utilizes TCP-based schemes to traverse NAT and establish reliable connections [5][6]. However, the inherent three-way handshake process of TCP introduces latency and lacks integrated TLS encryption, rendering TCP-based hole punching schemes inadequate for the growing demand for fast and secure communication.

In 2012, Google introduced a QUIC protocol [7], which was standardized by the Internet Engineering Task Force (IETF) in 2021 [8]. QUIC achieves low-latency, reliable, and secure network connections [9], showing potential to overcome the bottlenecks associated with TCP-based hole punching. However, there is a lack of research on the application of the QUIC protocol in NAT hole punching scenarios. While studies have analyzed the performance [10] and security [11] of the QUIC protocol, and compared it with TCP [10],[12] these investigations have not focused on performance disparities in NAT hole punching scenarios. M. Seemann *et al*. compared hole punching success rates of QUIC and TCP in NAT hole punching scenarios, revealing QUIC's superiority in hole punching. However, the study did not delve into a comparison of hole punching time overheads [13].

To fill the aforementioned research gap, we present a hole punching scheme for NAT traversal based on the standardized QUIC protocol. By configuring various network metrics, we compared hole punching time between QUIC-based and TCP-based protocols, revealing more efficiency of QUIC-based hole punching. Additionally, in scenarios where hole punching connections are disrupted due to network switching or NAT timeouts, this paper further explores a QUIC connection migration scheme for restoring connectivity. Our evaluation shows that this scheme saves 2 RTTs compared to QUIC re-punching and 3 RTTs compared to TCP re-punching. The results

This work is supported in part by the Guangdong Basic and Applied Basic Research Foundation (2024A1515012407), and in part by the Shenzhen Science and Technology Program (JCYJ20210324094609027).

highlight the advantages of QUIC in adapting to dynamic network environments.

This paper is structured as follows: Section II provides background information about NAT mapping rules, hole punching process, and a comparison of QUIC-based and TCP-based hole punching schemes. Section III focuses on the experimental deployment scheme and analysis of results. Section IV explores the potential advantages of QUIC connection migration for NAT hole punching. Finally, Section V summarizes our work.

## II. BACKGROUND

This section introduces NAT mapping rules, the hole punching process, and a comparison between QUIC-based and TCP-based hole punching schemes.

### A. NAT Mapping Rules

NAT technology can be classified into two types of mapping rules [14].

*1) Endpoint-Independent Mapping:* A private node $P$ is mapped to a public address $nodeP$ by a NAT device, and the mapping remains consistent regardless of connecting to different public nodes. When communicating with public nodes $N1$ or $N2$, $P$ establishes connections using $nodeP$, as shown in Fig. 1a.

*2) Address and Port-Dependent Mapping:* A private node $P$ is mapped to a public address $nodeP$ by a NAT device, and the mapping is dynamically adjusted with different public nodes. When communicating with public node $N1$, $P$ is mapped to public address $nodeP1$ by the NAT; when communicating with $N2$, $P$ is mapped to $nodeP2$ by the NAT, as shown in Fig. 1b.

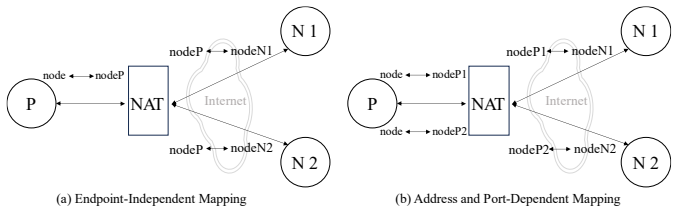

Fig. 1. NAT mapping rules

If the Endpoint-Independent Mapping rule is adopted, once $N1$ establishes a connection with $P$, it can forward $nodeP$ to $N2$ or other nodes, facilitating NAT hole punching. If the Address and Port-Dependent Mapping rule is adopted, $P$ uses $nodeP1$ to connect with $N1$, but is unable to use $nodeP1$ to connect with $N2$ or other nodes, hindering NAT hole punching.

### B. NAT Hole Punching Process

This study focuses on the most common NAT hole punching scenario, which involves communication between two end nodes located behind different NAT devices (shown in Fig. 2). In the following, we refer to ClientA, ClientB, and Relay Server as A, B, and S, respectively.

- First, both A and B establish a connection with S, respectively.
- Upon successful connection, A sends a registration message to S and opens a listening port. S records A's public address $(IP_A: PORT_A)$ and private address $(ip_A: port_A)$. Similarly, B sends a registration message to S and opens a listening port. S also records B's public address $(IP_B: PORT_B)$ and private address $(ip_B: port_B)$.

- Suppose A wishes to establish a connection with B, A requests B's public address from S. S sends B's public address to A and A's public address to B. At this point, both A and B know each other's public addresses (see Fig. 2a).

- A initiates a connection request to B. As the request passes through NAT-A, a session table entry is created on NAT-A, with the source address as $(ip_A: port_A)$ and the destination address as $(IP_B: PORT_B)$. Upon reaching NAT-B, the request is discarded as an unauthorized external connection because NAT-B lacks a matching session entry. Subsequently, B initiates a connection request to A, creating a session entry on NAT-B with the source address as $(ip_B: port_B)$ and the destination address as $(IP_A: PORT_A)$. Upon reaching NAT-A, the request is forwarded to A due to the existing session entry from A-to-B in NAT-A, thus opening the "hole" between A and B (see Fig. 2b).

- A and B begin to establish a connection. If the connection is successful, A and B can communicate directly, and subsequent data transmission does not require intermediation through S.

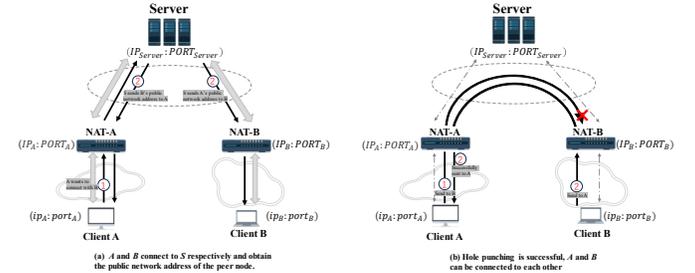

Fig. 2. NAT hole punching process

### C. Comparison of QUIC and TCP Hole Punching

Despite the similarities in the hole punching process for QUIC and TCP protocols, their implementations differ due to the inherent characteristics of each protocol.

*1) Port Multiplexing:* TCP sockets adhere to a one-to-one response mode, meaning that a local port can only be bound to one socket [15]. Therefore, TCP hole punching cannot establish simultaneous outbound and inbound connections on the same port, requiring port multiplexing. In contrast, QUIC permits multiple sockets to be bound to the same local port [16], eliminating the need to implement port multiplexing.

*2) Hole Punching Time:* Assuming negligible latency differences in network environments of clients $A$, $B$, and relay server $S$, according to the steps described in Section II-B, the hole punching time includes 1 RTT for clients to request peer's public addresses from $S$, at the time for $A$ and $B$ to establish a connection. A QUIC connection requires 1 RTT, a TCP connection requires 1.5 RTTs. Thus the hole punching time for QUIC is 2 RTTs, and for TCP, it is 2.5 RTTs.

*3) Connection Security:* QUIC, based on UDP and operating in user space, integrates with TLS 1.3 (see Fig. 3), enabling key negotiation, authentication, and session resumption within 1 RTT, thus providing higher security [9]. TCP operates in kernel space and cannot directly integrate with TLS 1.3. Thus, the connection processes of TCP and TLS 1.3 are independent, requiring at least 2 RTTs to establish a secure connection (The third handshake of a TCP connection can be sent simultaneously with the first handshake of a TLS 1.3 connection). This characteristic disadvantages TCP hole punching in terms of time efficiency and security.

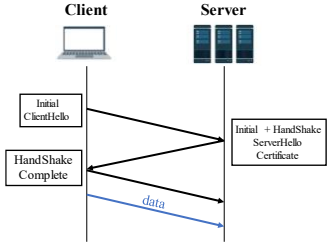

Fig. 3. Connection process of QUIC integrated with TLS 1.3

*4) Connection Restoration:* If hole punching is successful but the connection is disrupted due to network instability, TCP hole punching requires re-punching, resulting in additional computational resource consumption and latency. While, QUIC connection migration provides a rapid restoration mechanism. Specifically, QUIC uses a five-tuple (source IP, destination IP, source port, destination port, and connection ID) for communication [9]. When the network environment of an end node changes, connection migration automatically transfer the existing connection to a new network path without interruption. This ensures the continuity of the connection during network switching. Further details will be discussed in Section IV.

We can summarize the bottlenecks of current TCP hole punching and the advantages of QUIC hole punching.

*1) Bottlenecks of TCP hole punching:*

*a)* Additional implementation of port multiplexing is required for coding.

*b)* While ensuring connection reliability, the TCP three-way handshake introduces additional latency.

*c)* TCP lacks encryption capabilities, and additional TLS connections are required for security.

*d)* After the connection is disrupted, besides re-punching, there are currently no better solutions available for swift connection restoration.

*2) Advantages of QUIC hole punching:*

*a)* Port multiplexing functionality can be implemented directly for coding.

*b)* QUIC, based on UDP and integrated with the TLS 1.3 protocol to enable faster connections and higher security.

*c)* The QUIC connection migration mechanism can rapidly restore network connectivity.

## III. EXPERIMENTAL STUDY OF QUIC HOLE PUNCHING SCHEME

Given the limitations of TCP in hole punching scenarios and the advantages of QUIC, we propose a QUIC-based hole punching scheme for NAT traversal. We conduct experimental research to validate its efficiency.

As described in Section II-A, the Address and Port-Dependent Mapping rule leads to dynamic changes in NAT mappings, complicating hole punching. Testing revealed that NAT devices in our current real network follow this rule. Therefore, we conduct experiments in a controlled environment using the Endpoint-Independent Mapping rule.

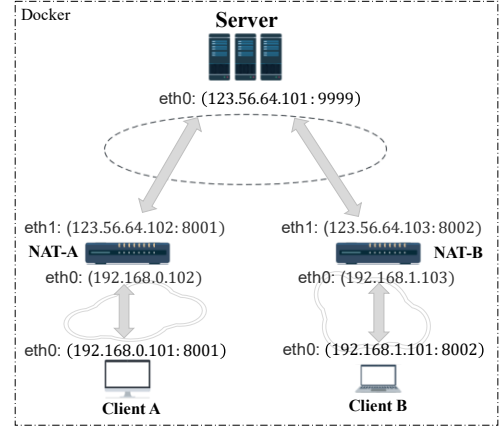

Fig. 4. Network environment configuration

### A. Experimental setup

To simulate a scenario where clients from two different LANs are behind different NAT devices, we set up the following network environment in a Docker container. ClientA is in subnet 192.168.0.0/24, and ClientB is in subnet 192.168.1.0/24. The relay server's address is 123.56.64.101:9999. The public addresses after mapping by NAT-A and NAT-B are 123.56.64.102:8001 and 123.56.64.103:8002, respectively (see Fig. 4). Additionally, we use iptables to configure the mapping rules of NAT-A and NAT-B, as detailed in TABLE I.

TABLE I. NAT MAPPING RULES

| NAT devices | Mapping Rules |
|---|---|
| NAT-A | iptables --append FORWARD --jump ACCEPT<br>iptables --append INPUT --jump DROP<br>iptables --append OUTPUT --jump DROP<br>iptables --table nat --append POSTROUTING --source 192.168.0.0/24 --jump SNAT --to-source 123.56.64.102 |
| NAT-B | iptables --append FORWARD --jump ACCEPT<br>iptables --append INPUT --jump DROP<br>iptables --append OUTPUT --jump DROP<br>iptables --table nat --append POSTROUTING --source 192.168.1.0/24 --jump SNAT --to-source 123.56.64.103 |

### B. Selection of Network Metrics

We first assess the impact of the following metrics on hole punching time, and then select appropriate combinations of network metrics.

*1) Round-Trip Time (RTT):* RTT is the time taken for a network request to travel from source to destination and back. The higher the RTT, the longer the hole punching time. Thus, RTT serves as a metric to measure hole punching performance. Based on literature [17], we selected three different RTT values of 20ms, 100ms, and 200ms for testing.

*2) Packet Loss Rate:* To assess the impact of packet loss rate on the experiment, we implemented a 1% packet loss rates within the Docker network. We observed that the hole punching time for both QUIC and TCP is significantly affected by packet loss, as shown in Fig. 5. Therefore, packet loss rate is a key performance metric.

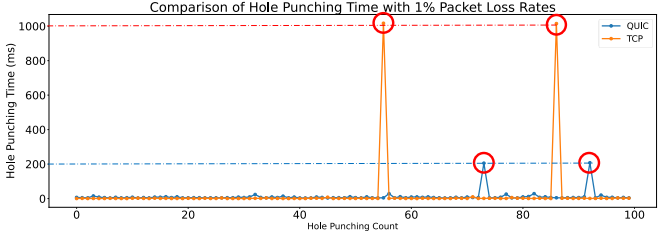

Fig. 5. QUIC and TCP hole punching time at 1% packet loss rates

In Fig. 5, we observe that the packet retransmission time for QUIC exceeds 200ms, while for TCP, it exceeds 1000ms. This discrepancy is due to their distinct retransmission mechanisms. According to draft-ietf-quic-recovery [18], QUIC's retransmission time is set to 200ms. Therefore, a packet loss on a QUIC connection results in a 200ms retransmission time. Similarly, RFC6289 [19] defines standard retransmission algorithms for TCP, as in (1).

$$RTO = \begin{cases} 1s, & RTO < 1s \\ RTO, & otherwise \end{cases} \quad (1)$$

Where $RTO$ is the retransmission timeout and the detailed calculation process can refer to RFC6289 [19]. The RTO is rounded up to 1s if it is less than 1s; otherwise, it remains unchanged. In our experiment, with a maximum RTT of 200ms, the $RTO$ does not exceed 1s. Consequently, TCP's retransmission timeout is calculated as 1s.

Based on literature [17], the Internet packet loss rate ranges between 1% and 2%. Thus, we set four different packet loss rates for testing: 0%, 1%, 1.5%, and 2%.

*3) Bandwidth:* Bandwidth determines the volume of data that can be transmitted in network communication. Although it is crucial in network evaluation, QUIC and TCP hole punching do not involve transmitting large amounts of data. Thus, bandwidth has little impact on the experimental results. To verify this, we conducted experiments in the Docker environment under four bandwidth conditions: unlimited, 10Gbps, 100Mbps, and 1Mbps. For each condition, we conducted 100 hole punching experiments and recorded the average time. Fig. 6 shows that there is minimal variation in hole punching time for QUIC and TCP across different bandwidths. Therefore, bandwidth is not considered a significant factor in our study.

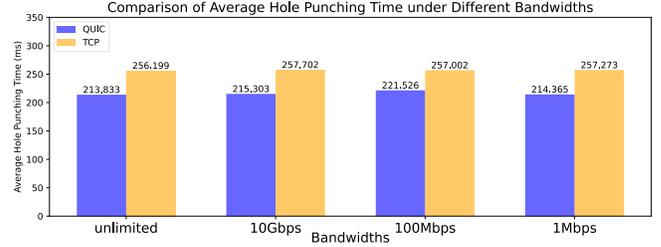

Fig. 6. The average hole punching time for QUIC and TCP under different bandwidths.

In summary, after analyzing the impact of various performance metrics on our experiment, we selected RTT and packet loss rate as the primary indicators. We tested 12 combinations, as detailed in TABLE II.

TABLE II. COMBINATION OF NETWORK SETTINGS FOR RTT AND PACKET LOSS RATE

| Experimental Combination | RTT（ms） | Packet Loss Rate |
|---|---|---|
| *1* | 20 | 0 % |
| *2* | 20 | 1 % |
| *3* | 20 | 1.5 % |
| *4* | 20 | 2 % |
| *5* | 100 | 0 % |
| *6* | 100 | 1 % |
| *7* | 100 | 1.5 % |
| *8* | 100 | 2 % |
| *9* | 200 | 0 % |
| *10* | 200 | 1 % |
| *11* | 200 | 1.5 % |
| *12* | 200 | 2 % |

*C. Experimental Testing*

This section outlines the process of experimental testing aimed at evaluating the hole punching time of QUIC and TCP.

As analyzed in Section II-C2, the ideal hole punching time is 2 RTTs for QUIC and 2.5 RTTs for TCP. However, we also need to consider the scenarios described in Steps 4 and 5 of Section II-B. With the presence of NAT devices, the first connection initiated by either client A or B will be discarded; only the second connection can pass through. It is necessary to analyze the time taken for the first connection to reach the peer's NAT device and the second connection to reach the local NAT device. Assuming the first connection initiated by A is $ConnA$, and the second connection initiated by B is $ConnB$.

*1) Scenario 1:* If $ConnA$ reaches NAT-B simultaneously with $ConnB$, ConnA can directly traverse NAT-B, establishing the first handshake. In this scenario, the hole punching time is 2 RTTs for QUIC and 2.5 RTTs for TCP (see Fig. 7a).

*2) Scenario 2:* If $ConnA$ reaches NAT-B before $ConnB$, NAT-B will discard $ConnA$, causing an additional delay of 0.5 RTTs. When $ConnB$ arrives at NAT-A, it becomes the first handshake connection as NAT-A has a session entry for $ConnB$. In this scenario, the hole punching time is 2.5 RTTs for QUIC and 3 RTTs for TCP (see Fig. 7b).

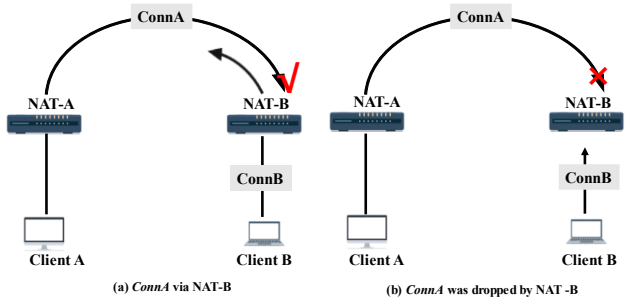

Fig. 7. Two extreme situations that occur during the hole punching process

In summary, we estimate the hole punching time to be between 2 and 2.5 RTTs for QUIC, and between 2.5 and 3 RTTs for TCP.

To validate the accuracy of our evaluation, we define the start time of the experiment as the moment when A initiates a request to S. The end time is when the connection between A and B is established. The hole punching time is calculated as the difference between these two timestamps. To ensure reliable results, we conduct 100 tests for each of the 12 combinations listed in TABLE II. We utilize the network simulation tool (*Netem*) on a Linux system and apply Traffic Control (TC) commands to mimic different RTTs and packet loss rates.

*D. Experimental Results*

This section analyzes the experimental results to validate our evaluation. Fig. 8 illustrates the hole punching time for QUIC and TCP under various network conditions.

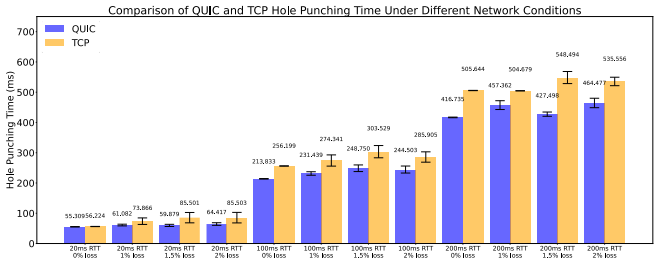

Fig. 8. Summary of QUIC and TCP hole punching time

Firstly, verifying under conditions without packet loss. With an RTT of 20ms, the hole punching time for QUIC is approximately 55ms, slightly higher than the expected 2 to 2.5 RTTs. This is because QUIC operates in user space, leading to some processing overhead even in a low-latency environment. Therefore, the observed hole punching time of approximately 55ms for QUIC still falls within the expected range. The hole punching time for TCP is approximately 56ms, within the expected 2.5 to 3 RTTs. When the RTTs are 100ms and 200ms, the hole punching time for QUIC is approximately 213ms and 416ms, respectively, within the expected range. The hole punching time for TCP is approximately 256ms and 505ms, respectively, also within the expected range.

Secondly, in the presence of packet loss, the fluctuation range of hole punching time for QUIC and TCP is similar to that without packet loss, indicating that they still perform as expected in such conditions.

The experimental results demonstrate that the QUIC-based hole punching scheme outperforms the TCP-based scheme in terms of hole punching time, especially in weak network environments. We believe that as the QUIC protocol becomes more widespread and adopted, its advantages in NAT hole punching scenarios will be further highlighted.

## IV. DISCUSSION

After completing NAT hole punching, connection maintenance remains a concern. In practical applications, public IP addresses of end nodes may change due to NAT session timeouts, NAT device reboots, or network transitions, disrupting the hole punched connection. Currently, the most common solution is re-punching [5], but this consumes additional computational resources. Therefore, finding a more effective method to maintain peer connections is crucial.

We evaluate a scheme to restore hole punched connections by utilizing the QUIC connection migration mechanism, and discussing its benefits in resolving disruptions in hole punched connections.

*A. QUIC Connection Migration Mechanism*

Connection migration implementation relies on the Connection ID (CID) in the QUIC protocol [8]. QUIC identifies a connection using a unique set of CIDs, binding the connection's state information to the CID rather than the underlying network address. As long as the CID remains unchanged, the connection persists even if the network address changes, and the upper-layer business logic remains unaware, eliminating the need for reconnection. This CID-based connection identification mechanism provides flexibility and portability to the QUIC protocol, allowing seamless migration of connections and maintaining communication continuity in changing network environments.

The process of connection migration is shown in Fig. 9:

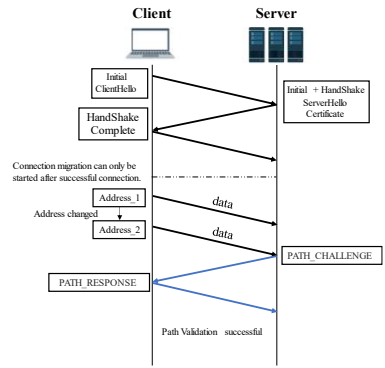

Fig. 9. Connection migration process

- When a client's address (IP and port) changes, it sends data to a server using a new address.

- The server receives the data and detects the change in the client's address. It then selects a new Connection ID (Destination Connection ID, DCID) from the previously maintained CID pool and sends a $PATH_C HALLENGE$ frame ($PC$ frame) to verify the reachability of the new address.

- If the new address is reachable, the client immediately responds with a $PATH_RESPONSE$ frame ($PR$ frame), confirming the reachability of the current path, and communication continues uninterrupted.

*B. Schemes for Restoring Connectivity*

This section analyzes and compares the processes and time overheads of two schemes for restoring connectivity: QUIC connection migration and re-punching. In the following discussion, we illustrate the case of client $A'$s private IP address changing.

*1) QUIC Connection Migration:* When the IP address of client $A$ changes, its public address also changes, disrupting the established punched connection. If A directly initiates a connection migration request to B, NAT-B will discard the request, preventing it from reaching B. To restore communication, the relay server S must be used for information exchange, as shown in Fig. 10a.

- A initiates a connection migration request to S, updating its own information, and requests S to send A's new public address to B. At the same time, A requests B to send data to trigger session establishment on NAT-B.
- S processes A's request and forwards A's new public address to B.
- Upon receiving S's message, B sends data to A. Since there is no corresponding session entry on NAT-A yet, this data is discarded. However, NAT-B records the session entry for data transmission from B to A, preparing for subsequent connection establishment.
- Because B's IP address and public address have not changed, A can directly send a connection migration request to B. Since there is already a session entry on NAT-B for data transmission from B to A, A's connection migration request can successfully pass through NAT-B and reach B, achieving connection migration and data transmission. Meanwhile, this connection request already has the capability to send data.

The total time for the above steps can be expressed as in:

$$T_{migrate} = T_{A2S} + T_{S2B} + T_{B2A} + T_{A2B} \quad (2)$$

Where $T_{migrate}$ is the time to restore connectivity, $T_{A2S}$ is the time for A to initiate connection migration and updates information to S, $T_{S2B}$ is the time for S processes the request and forwards it to B, $T_{B2A}$ is the time for B sends data to A, $T_{A2B}$ is the time for A sends a connection migration request to B.

*2) Re-punching with QUIC and TCP:* In the absence of connection migration functionality in QUIC or TCP, when the IP address of $A$ changes, the connections between $A$ and $B$, and between $A$ and $S$, are disrupted. If $A$ directly initiates a connection to $B$, this message will be discarded by NAT-B and cannot reach $B$. An effective solution is re-punching, as illustrated in Fig. 10b.

- A reestablishes QUIC/TCP connection with S.
- After establishing the connection, A sends a data request to S, asking S to send A's new public address to B and requesting B to send data to A.
- S processes A's request and forwards A's new public address to B.
- Upon receiving the message from S, B sends data to A. Since there is no corresponding session entry on NAT-A yet, the data sent by S to A is discarded. However, NAT-B records the session entry for data transmission from B to A.
- Since B's IP address and public address remain unchanged, A's request to establish a connection can successfully pass through NAT-B to reach B.
- Once the connections are successfully established, data transmission can occur.

The total time for the above steps can be expressed as in:

$$T_{re-punching} = T\grave{}_{A_S} + T\grave{}_{A2S} + T\grave{}_{S2B} + T\grave{}_{B2A} + T\grave{}_{A_B} + T\grave{}_{A2B} \quad (3)$$

Where $T_{re-punching}$ is the time for re-punching; $T\grave{}_{A_S}$ is the time for A and S to establish a new QUIC/TCP connection; $T\grave{}_{S2B}$ is the time for A sends data to S; $T\grave{}_{S2B}$ is the time for S processes the message and communicates it to B; $T\grave{}_{B2A}$ is the time for B sends a message to A's new public address; $T\grave{}_{A_B}$ is the time for A initiates the re-establishment of the QUIC/TCP connection to B; $T\grave{}_{A2B}$ is the time for A receives the message from the peer after successful establishment.

*3) Comparative Analysis:* We can observe similarities between Section IV-B1 and IV-B2, assuming that the network environments for both schemes are the same. The main difference lies in whether it is necessary to re-establish the QUIC/TCP connections between $A$ and $S$, and $A$ and $B$. TABLE III. compares the steps and time of the two schemes.

TABLE III.  COMPARES THE STEPS AND TIME OF THE TWO SCHEMES

| Processes | QUIC connection migration (steps : time) | QUIC/TCP Re-punching (steps : time) |
| --- | --- | --- |
| Information exchange | Steps 1-3 : $T_{A2S} + T_{S2B} + T_{B2A}$ | Steps 2-4 : $T\grave{}_{A2S} + T\grave{}_{S2B} + T\grave{}_{B2A}$ |
| Data interaction between $A$ and $B$ | Step 4 : $T_{A2B}$ | Step 6 : $T\grave{}_{A2B}$ |
| Establish connections between $A$ and $S$, and $A$ and $B$ | --- | Steps 1, 5 : $T\grave{}_{A_S} + T\grave{}_{A_B}$ |

Combining equations (2) and (3), we have:

$$\Delta t = T_{re-punching} - T_{migrate} = T\grave{}_{A_S} + T\grave{}_{A_B} \quad (4)$$

Where $\Delta t$ represents the time difference between re-punching and connection migration. It can be concluded that the time for connection migration to restore connectivity is reduced by 2 RTTs compared to QUIC re-punching and by 3 RTTs compared to TCP re-punching, further highlighting the advantage of QUIC in NAT hole punching scenarios.

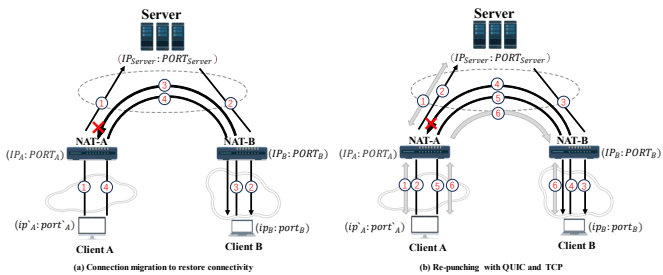

Fig. 10. Two options for restoring connectivity

## V. CONCLUSION

This paper presents and evaluates a QUIC-based hole punching scheme for NAT traversal, contrasting it with a TCP-based hole punching scheme. Our experiments show that QUIC hole punching reduces connection time by optimizing the connection process and performs better in weak network environments. Additionally, we further explore the application of the QUIC connection migration mechanism in handling changes in end node addresses. Our evaluation results indicate restoring connectivity through QUIC connection migration saves 2 RTTs compared to QUIC re-punching and 3 RTTs compared to TCP re-punching, thus reducing the computational resources required for reconnecting.

In summary, the QUIC-based hole punching scheme for NAT traversal exhibits certain advantages in efficiency and security. This scheme provides a new direction for NAT hole punching technology and is expected to play an important role in future network communication.